\documentclass[twocolumn,showpacs,preprintnumbers,amsmath,amssymb]{revtex4}

\usepackage{graphicx}% Include figure files
\usepackage{dcolumn}% Align table columns on decimal point
\usepackage{bm}% bold math
\usepackage{mathrsfs}
\usepackage{color}

\begin{document}

\title{Improved version of simplified method for
including tensor effect in cluster models
}% 

\author{N. Itagaki$^1$ and A Tohsaki$^2$}

\affiliation{
$^1$Yukawa Institute for Theoretical Physics, Kyoto University,
Kitashirakawa Oiwake-Cho, Kyoto 606-8502, Japan
}

\affiliation{
$^2$Research Center for Nuclear Physics (RCNP), Osaka University,
10-1 Mihogaoka, Ibaraki, Osaka 567-0047, Japan
}

\date{\today}

\begin{abstract}
The tensor contribution can be directly incorporated in the cluster model
in a simplified way.
In conventional $\alpha$ cluster models, the contribution of the
non-central interactions exactly cancels because of the antisymmetrization effect
and spatial symmetry of $\alpha$ clusters.
The mixing of breaking components
of $\alpha$ clusters to take into account the spin-orbit and tensor effects is needed.
Previously we proposed a simplified method to include the spin-orbit effect,
and also for the tensor part,
a simplified model to directly take into account the contribution of the tensor interaction
(called SMT) was introduced;
however the contribution of the tensor interaction was quite limited.
Here we improve SMT, which is called $i$SMT.
Using newly proposed $i$SMT, the contribution  of the tensor interaction 
in $^4$He is more than $-40$ MeV, four times larger than the previous version.
The method is applied to four-$\alpha$ cluster structure of $^{16}$O.
In $^{16}$O, the tensor contribution is also large,
and this is coming from the finite size effect for the distances among 
$\alpha$ clusters with a tetrahedral configuration.
\end{abstract}

\maketitle

\section{Introduction}
The binding energy per nucleon of $^4$He is quite large in light mass region and $\alpha$ particles 
are considered as good building blocks for the nuclear structure.
Cluster models, especially the $\alpha$ cluster models, are based on this idea, and  
they have been widely used for the description of molecular structure
of nuclei~\cite{Brink,Fujiwara}.
One of the well-known examples is the so-called Hoyle state~\cite{Hoyle};
formation of  $^{12}$C from three $^4$He nuclei ($\alpha$ clusters) is a key process
of the nucleosynthesis. The second $0^+$ state
at $E_x = 7.6542$ MeV plays a crucial role, which is 
the second excited state of $^{12}$C and located
just above the threshold energy  to decay into three $^4$He nuclei.
The existence of a state which has the character of three $\alpha$ clusters just at this energy is
really an essential factor in the synthesis of various elements in stars.
Such three-$\alpha$-state is described by various cluster models,
and among them, the Tohsaki-Horiuchi-Schock-R\"{o}epke (THSR) wave function is a powerful
tool to describe gas-like cluster states with spatial extension~\cite{THSR}.
Based on the shell-model picture, which is standard in nuclear structure physics,
we often need large model space to 
describe cluster states. Since some of the nucleons are spatially correlated around 
the nuclear surface, the cluster states are difficult to be described
with a framework in which the wave function of each nucleon is expanded
around the origin. 
Therefore the cluster structures are challenge of the shell models, including modern $ab\ initio$ 
ones~\cite{Maris,Dreyfuss,Yoshida}.

Nuclear systems have characteristic features that
non-central interactions play a crucial role; however
in most of the cluster models, 
the spin-orbit and tensor interactions
do not contribute inside $\alpha$ clusters and also between $\alpha$ clusters
because of the antisymmetrization effect
and spatial symmetry of $\alpha$ cluster.
In cluster models, each $\alpha$ cluster is often defined as 
a simple $(0s)^4$ configuration at some spatial point, and
$\alpha$ cluster is a spin singlet system, which is free from
the non-central interactions.
Concerning the spin-orbit interaction, this is known to be 
quite important in explaining the observed magic numbers.
The $jj$-coupling shell model, which is the standard model for the nuclear structure,
is based on this picture. 

Our goal is to pave the way to generally describe the nuclear structure, including shell and cluster structures
simultaneously.
Here, contrary to the standard approaches, we start with the cluster model side and try to include shell correlations.
This is because our approach requires much less computational efforts compared with the case
starting with the shell model side.
To include the spin-orbit contribution
starting with the cluster model,
we proposed the antisymmetrized quasi-cluster model 
(AQCM)~\cite{Simple,Masui,Yoshida2,Ne-Mg,Suhara,Suhara2015,Itagaki,Itagaki-CO,Matsuno},
which allows smooth transition of $\alpha$ cluster model wave function to
$jj$-coupling shell model one.
In AQCM, this transition 
can be controlled by only two parameters: $R$ representing the distance between $\alpha$ clusters
and $\Lambda$, which characterizes the transition of $\alpha$ cluster(s) to 
$jj$-coupling shell model wave functions
and quantifies the role of the spin-orbit interaction. 
We call the transformed $\alpha$ clusters in this way quasi-clusters.
As it is well known,
the conventional $\alpha$ cluster models cover the model space of closure of major shells
($N=2$, $N=8$, $N=20$, {\it etc.}) and in addition, 
the subclosure configurations of the $jj$-coupling shell model,
$p_{3/2}$ ($N=6$), $d_{5/2}$ ($N=14$), $f_{7/2}$ ($N=28$), and $g_{9/2}$ ($N=50$)
can be described by our AQCM. 
In this way the cluster and $jj$-coupling shell model wave functions can be described on the same footing,
and the spin-orbit interaction, which is the rank one non-central interaction,
can be successfully taken into account in the cluster model. 

However the rank two non-central
interaction, the tensor interaction, is more complicated to be treated in the cluster model.
The tensor interaction has two features, the first order type and the second order type.
The first order one is rather weak and 
characterized by the attractive effect for a proton (neutron) with the 
$j$-upper orbit of the $jj$-coupling shell model and a neutron (proton) with $j$-lower orbit, 
or repulsive effect for the $j$-upper ($j$-lower) two protons 
or two neutrons~\cite{Otsuka}.
This effect can be included just by switching on the tensor interaction
in the Hamiltonian, after transforming cluster wave function to
 $jj$-coupling shell model one using AQCM mentioned above.
The second order effect of the tensor is much stronger.
According to the {\it ab initio} calculations, the (negative) contribution of the tensor interaction
in $^4$He is quite large, more than $-65$ MeV~\cite{Kamada}, 
and this is even more important than the central interaction.
Here, it is found that the 
two particle two hole (2p2h) excitation to higher shells, especially to the $p$ shell, is quite important.
According to the tensor optimized shell model (TOSM) 
calculations~\cite{TOSM,TOSM2007,TOSM2009,TOSM2011,TOSM2012},
the $p$ orbits of this 2p2h states
must have very shrunk shape compared with the normal shell model orbits, 
and this means that mixing of very high momentum components 
is quite important. 

This second order effect of the tensor interaction is more difficult to be treated
in the cluster model, and we need an additional framework; 
we have proposed 
a simplified model to directly take into account the contribution of the tensor interaction
(SMT)~\cite{Itagaki-SMT}. 
The tensor contribution was estimated in $^4$He, $^8$Be, and $^{12}$C,
and the relation to the clustering was quantitatively discussed. 
However the contribution of the tensor interaction was rather limited, about $-10$ MeV in
the $\alpha$ cluster, and improvement of the model was needed. 
In our previous SMT, we started with an $\alpha$ cluster with a $(0s)^4$ configuration and 
expressed deuteron-like excitation of a proton and neutron to higher shells
by shifting the values of the Gaussian center parameters 
of these two particles.
However, shifting the positions of Gaussian center parameters may not be enough
for the purpose of mixing higher momentum components of 2p2h configurations,
and this could be the reason.
In the present article, we introduce improved version of SMT, which is $i$SMT.
Here, imaginary part of Gaussian center parameters is shifted in stead of the real part.
The imaginary part of Gaussian center parameter corresponds to the expectation value
of momentum for the nucleon.
The tensor interaction has the character which is suited to be described in the momentum space,
and this method is considered to be more efficient in directly mixing the
higher momentum components of 2p2h configurations.

The purpose of the present work is to incorporate the 2p2h nature 
of the tensor contribution 
in the cluster model
in a simplified and more efficient way compared with the previous SMT.
We improve SMT and newly propose $i$SMT.
Firstly we apply it to $^4$He and next discuss that the clustering of four $\alpha$'s
is closely related to the tensor effect in $^{16}$O.
There have been fundamental
discussion for the appearance of cluster structure in the
1960s; 
one-pion exchange potential (OPEP) vanishes in the direct
terms when each $\alpha$ cluster is described as a $(0s)^4$ configuration,
and this is the reason why inter-cluster interaction is weak.
However, it is important to show that
clustering is
still important, even if the model space is extended and the 
tensor contributions in each $\alpha$ cluster is taken into account. 
We discuss that the clustering is enhanced 
because of the tensor interaction in $^{16}$O.

Recently, many other attempts of directly taking
into account the tensor part of the interaction in microscopic
cluster models have begun.
For instance, by combining unitary correlation method (UCOM) and 
Fermionic molecular dynamics (FMD)~\cite{Neff,Roth,Chernykh},
or using antisymmetrized molecular dynamics (AMD)~\cite{Dote}, 
cluster structure has been extensively studied.
In UCOM, the tensor contribution can be taken into account
by  unitary transforming the Hamiltonian, where two-body correlator 
is introduced in the exponent of the unitary operator.
If we expand this power based on the cluster expansion method,
in principle, the Hamiltonian contains may-body operators up to $A$ (mass number) body,
thus the truncation of the model space is required.
Our strategy is slightly different. Although the framework is phenomenological,
we do not perform the unitary transformation of 
the Hamiltonian, and we introduce an effective model wave function
to directly take into account the tensor effect.

For the central part of the interaction, 
we use the Tohsaki interaction, which has finite range three-body terms.
This interaction is a phenomenological one and 
designed to reproduce the $\alpha$-$\alpha$ scattering phase shift.
Also it gives reasonable size and binding energy of the $\alpha$ cluster,
which is rather difficult in the case of the zero-range three-body interaction, and
the binding energy is less sensitive to the choice of size parameter of Gaussian-type
single particle wave function.
Furthermore, the saturation property is reproduced rather satisfactory.

% The paper is organized as follows. 
% The formulation is given in Sect.~\ref{model}. 
% In Sect.~\ref{results}, the AQCM results are shown. 
% Finally, in Sect.~\ref{summary} we summarize the results and give the main conclusions.

\section{The Model\label{model}}

\subsection{Hamiltonian}

The Hamiltonian ($\hat{H}$) consists of kinetic energy ($\hat{T}$) and 
potential energy ($\hat{V}$) terms,
\begin{equation}
\hat{H} = \hat{T} +\hat{V},
\end{equation}
and the kinetic energy term is described as one-body operator,
\begin{equation}
\hat{T} = \sum_i \hat{t_i} - T_{cm},
\end{equation}
and the center of mass kinetic energy ($T_{cm}$),
which is constant,
is subtracted.
The potential energy has
central ($\hat{V}_{central}$), spin-orbit ($\hat{V}_{spin-orbit}$), tensor ($\hat{V}_{tensor}$), 
and the Coulomb parts.

For the central part of the potential energy
($\hat{V}_{central}$), the Tohsaki interaction is adopted,
which consists of two-body ($V^{(2)}$)  and three-body ($V^{(3)}$) terms:
\begin{equation}
\hat{V}_{central} = {1 \over 2} \sum_{i \neq j} V^{(2)}_{ij} 
+ {1 \over 6} \sum_{i \neq j, j \neq k, i \neq k}  V^{(3)}_{ijk},
\end{equation}
where $V^{(2)}_{ij}$ and $V^{(3)}_{ijk}$ consist of three terms with different range parameters,
\begin{equation}
V^{(2)}_{ij} =  
\sum_{\alpha=1}^3 V^{(2)}_\alpha \exp[- (\vec r_i - \vec r_j )^2 / \mu_\alpha^2]
 (W^{(2)}_\alpha + M^{(2)}_\alpha P^r)_{ij},
\label{2body}
\end{equation} 
\begin{eqnarray}
V^{(3)}_{ijk} = 
\sum_{\alpha=1}^3 && V^{(3)}_\alpha \exp[- (\vec r_i - \vec r_j )^2 / \mu_\alpha^2 -
                                                     (\vec r_i - \vec r_k)^2 / \mu_\alpha^2   ] \nonumber \\
\times &&  (W_\alpha^{(3)} + M_\alpha^{(3)} P^r)_{ij} (W_\alpha^{(3)} + M_\alpha^{(3)} P^r)_{ik}.
\end{eqnarray}
Here, $P^r$ represents the exchange of spatial part of the wave functions
of interacting two nucleons.
In this article, we use
F1' parameter set~\cite{Itagaki-CO}, which was designed to avoid small overbinding of $^{16}$O
when the original F1 parameter set is adopted.
The difference of F1 and F1' is only for the three-body Majorana exchange parameter
for the shortest range.

For the spin-orbit part,
G3RS \cite{G3RS}, which is a realistic
interaction originally determined to reproduce the nucleon-nucleon scattering phase shift, 
is adopted;
\begin{equation}
\hat{V}_{spin-orbit}= {1 \over 2} \sum_{i \ne j} V^{ls}_{ij},
\end{equation}
\begin{equation}
V^{ls}_{ij}= V_{ls}( e^{-d_{1} (\vec r_i - \vec r_j)^{2}}
                    -e^{-d_{2} (\vec r_i - \vec r_j)^{2}}) 
                     P(^{3}O){\vec{L}}\cdot{\vec{S}}.
\label{Vls}
\end{equation}
For the strength,
$V_{ls} = 1800$ MeV has been suggested to reproduce the
various properties of $^{12}$C~\cite{Itagaki-CO}, and we use this value.

Up to this point, the interaction is the same as in Ref.~\cite{Itagaki-CO},
and the main purpose of the present article is to switch on the tensor interaction.
For the tensor part, we use Furutani interaction~\cite{Furutani}.
This interaction nicely reproduces the tail region of one pion exchange potential,
and the comparison is shown as Fig.~1 in Ref.~\cite{Itagaki-SMT}.

\subsection{Wave function }

The single particle wave function has a Gaussian shape \cite{Brink};
\begin{equation}
	\phi_{i} = \left( \frac{2\nu}{\pi} \right)^{\frac{3}{4}}
	\exp \left[- \nu \left(\bm{r}_{i} - \bm{R}_i \right)^{2} \right] \eta_{i},
\label{Brink-wf}
\end{equation}
where $\eta_{i}$ represents the spin-isospin part of the wave function, 
and $\bm{R}_i$ is a parameter representing the center of a Gaussian 
wave function for the $i$-th particle.
The size parameter $\nu$ is chosen to be 
0.25 fm$^{-2}$ for $^4$He (0.20 fm$^{-2}$ for $^{16}$O).
In Brink-Bloch wave function, 
four nucleons in one $\alpha$ cluster share a common and real value for the Gaussian center parameter. 
Hence, the contribution of the spin-orbit and tensor interactions vanishes.

The wave function of the total system $\Psi$ is antisymmetrized product of these
single particle wave functions;
\begin{equation}
\Psi = {\cal A} \{ \psi_1 \psi_2 \psi_3 \cdot \cdot \cdot \cdot \psi_A \},
\label{total-wf}
\end{equation}  
where $A$ is a mass number.
The projections onto parity and angular momentum eigenstates can be performed by 
introducing the projection operators $P^J_{MK}$ and $P^\pi$,
and these are performed
numerically in the actual calculation.

Based on generator coordinate method (GCM), the superposition of
different Slater determinants can be done,
\begin{equation}
\Phi = \sum_i c_i P^J_{MK} P^\pi \Psi_i.
\label{GCM}
\end{equation}
Here, $\{ \Psi_i\}$ is a set of Slater determinants,
and the coefficients for the linear combination, $\{ c_i \}$, are
obtained by solving the Hill-Wheeler equation~\cite{Brink}.

\subsection{SMT and $i$SMT}
Here we explain how we can incorporate the tensor effect starting with the $\alpha$ cluster model.
Previously, we have introduced 
SMT~\cite{Itagaki-SMT}. In SMT, we started with an $\alpha$ cluster with a $(0s)^4$ configuration
and changed it for the purpose of including the tensor contribution. 
For the $(0s)^4$ configuration,
the Gaussian center parameter ($\vec R$ in Eq.~\ref{Brink-wf}) 
for the spin-up proton ($\vec R_{p \uparrow}$),
spin-down proton ($\vec R_{p \downarrow}$),
spin-up neutron ($\vec R_{n \uparrow}$), and
spin-down neutron ($\vec R_{n \downarrow}$) were all set to zero.
In SMT, we mimicked deuterons, where a proton and a neutron have aligned spin orientation 
and spatially displaced in this spin orientation.
For $^4$He, we transformed it to two deuterons with spin up and down;
we shifted the Gaussian center of the spin-up proton to the $z$ direction, which forms a deuteron 
together with  the spin-up neutron at the origin, and we also shifted the spin-down neutron
to the $-z$ direction, which forms a deuteron with the spin-down proton at the origin.
The Gaussian center parameters were introduced in the following way;
\begin{eqnarray}
\vec R_{p \uparrow} &=& d {\vec e_z}, \\
\vec R_{n \uparrow} &=& 0, \\
\vec R_{p \downarrow} &=& 0, \\
\vec R_{n \downarrow} &=& -d {\vec e_z},
\label{SMT}
\end{eqnarray}
where $d$ is a distance parameter and $\vec e_z$ is a unit vector 
for the $z$ direction. We prepared Slater determinants with different 
$d$ values and superposed them based on GCM.
The adopted $d$ values in Ref.~\cite{Itagaki-SMT} were
0, 0.7, 1.4, 2.1, $\cdot \cdot \cdot$ 7.0 fm (11 Slater determinants in total).

\begin{figure}[t]
	\centering
	\includegraphics[width=6.0cm]{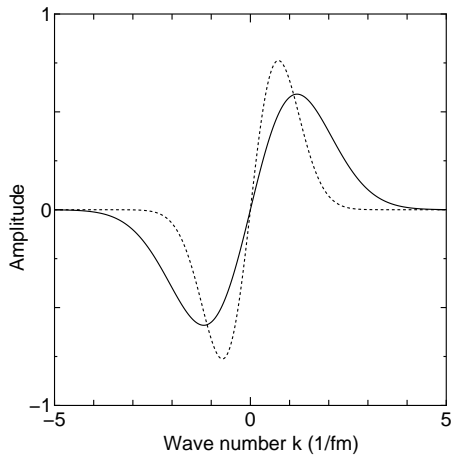} 
	\caption{
Fourier transformation of the $p$ orbit (one dimension).
The horizontal axis is the wave number $k$ (fm$^{-1}$).
The dotted line is is for the normal $p$ orbit with the standard size parameter ($b = 1.4$ fm),
and the solid line is the one
used in TOSM with a shrunk size parameter ($b = 0.6 \times 1.4$ fm).
     }
\label{ft}
\end{figure}

However, using this previous version of SMT,
the contribution of the tensor interaction was rather limited, about $-10$ MeV in
the $\alpha$ cluster. 
Shifting the positions of Gaussian center parameters may not be enough
for the purpose of mixing higher momentum components of 2p2h configurations,
and this could be the reason why the effect of the tensor interaction was rather limited.
According to TOSM, 
the higher-nodal orbits of the 2p2h states
must be introduced to have very shrunk shape compared with the normal shell model orbits.
In Fig.~\ref{ft}, Fourier transformation of one dimensional $p$ orbits are shown.
The $p$ orbit on the $x$ axis before the Fourier transformation
is proportional to $x\exp[-\nu x^2]$, and $\nu = 1/2b^2$.
The horizontal axis is the wave number $k$ (fm$^{-1}$).
The dotted line is for the normal $p$ orbit after the Fourier transformation 
with the standard size parameter ($b = 1.4$ fm),
and the solid line is the one 
used in TOSM (solid line) with a shrunk size parameter ($b = 0.6 \times 1.4 = 0.84$ fm).
After the Fourier transformation, the shrinkage in the coordinate space changes to 
the extension in the the momentum space;
the solid line is distributed in much larger $|k|$ region compared with the dotted line.
The root mean square of $k$ is 1.46 (fm$^{-1}$) and 0.87 (fm$^{-1}$) for the solid and dotted
line, respectively.

In the present article, we introduce improved version of SMT, which is $i$SMT.
Here, imaginary part of Gaussian center parameters is shifted in stead of the real part.
Using the single particle wave function (Eq.~\ref{Brink-wf}), we can show that the
expectation value of the momentum of the nucleon is proportional to the imaginary part
of the Gaussian  center parameter,
\begin{equation}
\langle {\vec p} \rangle = 2\nu \hbar Im ({\vec R}).
\label{momentum}
\end{equation}
In the present calculation, $\nu$ is chosen as 0.25 fm$^{-2}$ for $^4$He,
thus the imaginary part of the Gaussian center parameter
and the wave number have the relation of
\begin{equation}
\langle {\vec k} \rangle = 0.5 \times Im ({\vec R} ({\rm fm})) \ \ {\rm (fm^{-1})}.
\label{wavenumber}
\end{equation}

In $i$SMT, the Gaussian center parameters are introduced in the following way;
\begin{eqnarray}
\vec R_{p \uparrow} &=& d {i \vec e_z}, \nonumber \\
\vec R_{n \uparrow} &=& 0, \nonumber \\
\vec R_{p \downarrow} &=& \nonumber 0, \\
\vec R_{n \downarrow} &=& -d {i \vec e_z},
\label{$i$SMT-1}
\end{eqnarray}
and $d$ values are 0, 1, 2,  $\cdot \cdot \cdot$ 10 fm (11 Slater determinants).
In addition, we prepare the basis states, where neutron spin-up is shifted in stead of
neutron spin-down;
\begin{eqnarray}
\vec R_{p \uparrow} &=& d {i \vec e_z}, \nonumber \\
\vec R_{n \uparrow} &=& -d {i \vec e_z}, \nonumber \\
\vec R_{p \downarrow} &=& 0, \nonumber \\
\vec R_{n \downarrow} &=& 0,
\label{$i$SMT-2}
\end{eqnarray}
and $d$ values are 1, 2, $\cdot \cdot \cdot$ 10 fm (10 Slater determinants).
Eventually, we superpose these 21 Slater determinants in total based on GCM.
According to Eq.~\ref{wavenumber},
the expectation value of the momentum for $d = 10$ fm is 5 fm$^{-1}$.
\section{Results}

\subsection{$^4$He}

\begin{figure}[t]
	\centering
	\includegraphics[width=6.0cm]{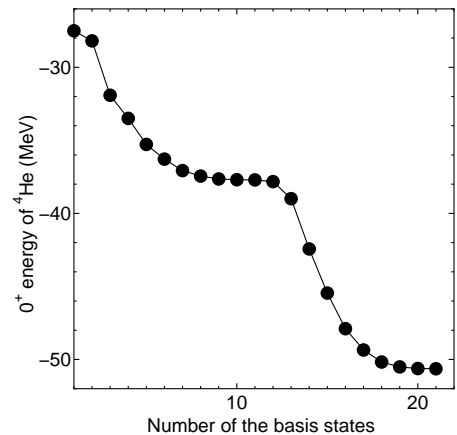} 
	\caption{
Energy convergence for the ground state 
of $^4$He described based on $i$SMT as a function of number of basis states.
     }
\label{he4-conv}
\end{figure}

First we start with $^4$He. 
The energy convergence for the ground state 
of $^4$He described based on $i$SMT 
is shown in Fig.~\ref{he4-conv}
as a function of number of basis states.
Here $i$SMT is a linear combination of 21 GCM basis states;
the basis state ``1" is $(0s)^4$ configuration,
and in 2-11 (12-21), the imaginary part of the Gaussian center parameters
for the spin-up proton and spin-down (spin-up) neutron are shifted as in Eq.~\ref{$i$SMT-1}
(Eq.~\ref{$i$SMT-2}), where $d$ values are 1, 2, 3, $\cdot \cdot \cdot$ 10 fm.
At ``1" on the horizontal axis, the tensor interaction does not contribute,
and the energy gets lower by more than 20 MeV with increasing the number of the GCM basis states.

In Table~\ref{compari}, we compare
the energies of $^4$He calculated using the $(0s)^4$ configuration,
conventional SMT, and newly introduced $i$SMT. Here total, 
$T$, $V^2$, $V^3$, $V^{ls}$, $V^t$,
and $V^{Coul}$ mean the expectation value of the total energy, kinetic energy, 
two-body interaction, three-body interaction,
spin-orbit interaction, tensor interaction, and Coulomb interaction, 
respectively.
The tensor contribution of $i$SMT is $-41.56$ MeV,
which is more than four times compared with the previous version.
The kinetic energy of $i$SMT increases 
from the value for the $(0s)^4$ configuration
by about 25 MeV in the positive direction, 
and this is because of the mixing of higher momentum components.
Here we can see that the contribution of the two-body interaction increases by about 9 MeV
in the negative direction.
Compared with the so called {\it ab initio} calculations,
the tensor contribution is still small, but we can include the effect
to the level of $-40$ MeV 
(our central part of the interaction is phenomenological one without
short range core, thus precise  comparison with {\it ab initio} calculations
is rather difficult).

\begin{table} 
 \caption{Energies of $^4$He calculated using the $(0s)^4$ configuration,
conventional SMT, and newly introduced $i$SMT. Here total, 
$T$, $V^2$, $V^3$, $V^{ls}$, $V^t$,
and $V^{Coul}$ mean the expectation value of the total energy, kinetic energy, 
two-body interaction, three-body interaction,
spin-orbit interaction, tensor interaction, and Coulomb interaction, 
respectively. All units are in MeV.}
  \begin{tabular}{cccc} \hline \hline
            & $(0s)^4$ & SMT & $i$SMT  \\ \hline
 total     & $-27.50$  &   $-32.85$   & $-50.64$ \\
 $T$      &    46.65    &    53.14   &  71.96 \\ 
 $V^2$   &  $-79.38$ & $-83.75$ &  $-88.65$\\
 $V^3$   &     4.41     &   6.22     & 6.18  \\
 $V^{ls}$ &   0.0         &  $0.11$   & $0.57$  \\
 $V^t$   &   0.0         &   $-9.40$ & $-41.56$  \\
 $V^{Coul}$ &   0.81   &   0.84     &  0.87 \\ \hline 
  \end{tabular}   \\ 
\label{compari}
\end{table}

\begin{figure}[t]
	\centering
	\includegraphics[width=6.0cm]{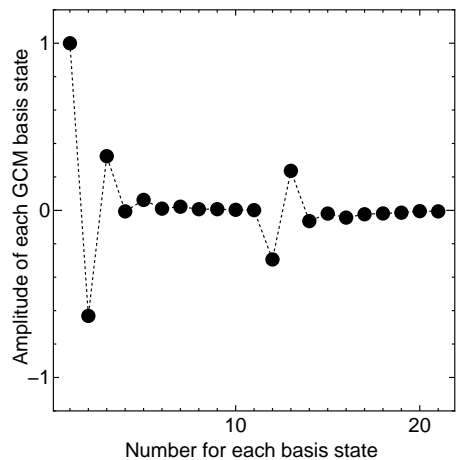} 
	\caption{
Amplitude for the liner combination of the basis states ($c_i$ in Eq.~\ref{GCM})
for $^4$He described based on $i$SMT. The amplitude for the
$(0s)^4$ configuration (basis state 1) is normalized to 1.
     }
\label{he4-amp}
\end{figure}

The amplitude for the liner combination of the basis states ($c_i$ in Eq.~\ref{GCM})
for $^4$He described based on $i$SMT is shown in Fig.~\ref{he4-amp}. 
In principle the amplitudes can be complex numbers; however here we obtained 
real numbers after diagonalizing the Hamiltonian. 
The basis state ``1" on the horizontal axis corresponds to the $(0s)^4$ configuration.
In the basis states 2-11 (12-21), the imaginary part of the Gaussian center parameters
for the spin-up proton and spin-down (spin-up) neutron are shifted as in Eq.~\ref{$i$SMT-1}
( Eq.~\ref{$i$SMT-2}).
The amplitude for the basis state ``2" is obtained as a negative value, and this is to create a node
for the wave function, which means particle-hole excitation (however we shift the Gaussian center parameters of
two particles simultaneously, thus direct correspondence to the $p$ orbits is rather difficult to be seen). The amplitude
for the basis state ``3" returns back to positive value, which means excitation to 
even higher shells, and absolute value of the amplitude gets smaller with increasing 
$d$ value. From ``12", spin-up neutron is shifted in stead of spin-down neutron.
Although the absolute value of the amplitude is smaller, the basis tendency is 
the same as the spin-down neutron case.

\subsection{$^{16}$O}

\begin{figure}[t]
	\centering
	\includegraphics[width=6.0cm]{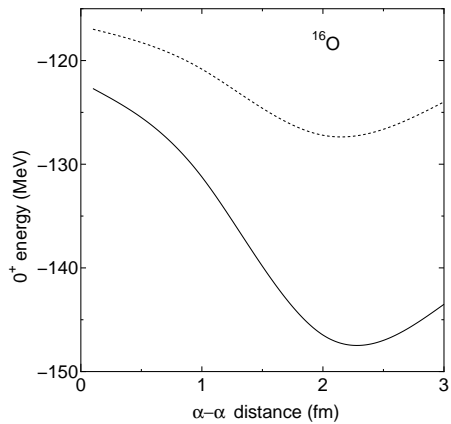} 
	\caption{
Energy curve for the $0^+$ state of $^{16}$O with a tetrahedral configuration of four-$\alpha$ clusters
as a function of distance between $\alpha$-$\alpha$. The solid and dotted line show
the results with and without the tensor interaction. 
     }
\label{O16-a}
\end{figure}

The same procedure can be applied to $^{16}$O and we can discuss the relation between
the tensor contribution and the clustering effect.
Here, $^{16}$O is introduced as a tetrahedral configuration of four-$\alpha$ clusters,
and one of the clusters is deformed using $i$SMT to include the tensor contribution.
The $0^+$ state energy of $^{16}$O with a tetrahedral configuration of four-$\alpha$ clusters
as a function of distance between $\alpha$-$\alpha$ is shown in Fig.~\ref{O16-a}. The solid and dotted lines show
the results with and without the tensor interaction. 
We can confirm that with the tensor interaction, the clustering is even enhanced.
The decrease of the energy after 
switching on the tensor interaction is only 5.7 MeV at the $\alpha$-$\alpha$ distance of 0.1 fm.
Here the contribution of the tensor interaction in the Hamiltonian is only $-10.2$ MeV.
The tensor contribution is suppressed 
at small relative distances (the wave function corresponds 
to the closed shell configuration of the $p$ shell at the zero-distance limit).
This is because, the 2p2h excitation from the lowest $s$ shell to the $p$ shell is forbidden,
even though the excitation from the $p$ shell to $sd$ shell is allowed. 
With increasing the $\alpha$-$\alpha$ distance, the decrease of the energy due to the tensor interaction
is enhanced. The decrease is about 20 MeV around the lowest energy point, and the matrix element of the 
tensor interaction is $-32.7$ MeV at the $\alpha$-$\alpha$ distance of 2 fm.
Therefore, it can be concluded that the tensor interaction has a certain effect for the stability of clusterized configurations.
In Refs.~\cite{Ogawa2006,Ogawa2011}, it has been suggested that tensor contribution is suppressed in $^{16}$O,
since 2p2h excitation from the lowest $s$ shell to the $p$ shell is forbidden at the shell model limit.
This is true; however tensor contribution turned out to be large in $^{16}$O because of the clustering effect 
of four $\alpha$'s.

In the present case, this 2p2h tensor effect is already renormalized in the central part
of the effective interaction as in many conventional models, 
and the result gives very large overbinding.  
In the next step, the modification of  the central part of the two-body and three-body interactions
to reproduce the binding energies of many nuclei including this kind of tensor effect will be carried out.

\section{Summary}\label{summary}

It has been shown that the tensor contribution can be incorporated in the cluster model
in a simplified way.
In conventional $\alpha$ cluster models, the contribution of the
non-central interactions cancels because of the antisymmetrization effect
and spatial symmetry of each $\alpha$ cluster,
and the mixing of the breaking components
of $\alpha$ clusters to take into account the spin-orbit and tensor effects is needed.
Previously we proposed a simplified method to include the spin-orbit effect,
and also for the tensor part,
a simplified method to take into account the tensor contribution 
in the cluster model (SMT) was introduced.
Here we improved SMT, which is called $i$SMT,
where the imaginary part of Gaussian center parameters of nucleons in one $\alpha$ cluster
was shifted in stead of the real part.
The imaginary part of Gaussian center parameter corresponds to the expectation value
of momentum for the nucleon.
The tensor interaction has the character which is suited to be described in the momentum space,
and this method is considered to be more efficient in directly mixing the
higher momentum components of 2p2h configurations.

Using newly proposed $i$SMT, the contribution  of the tensor interaction 
in $^4$He is more than $-40$ MeV, four times larger than the previous version.
The method was applied to four-$\alpha$-cluster structure of $^{16}$O.
In $^{16}$O,
the tensor contribution is suppressed 
at the limit of small relative distance corresponding to the closed shell configuration of the $p$ shell.
With increasing the $\alpha$-$\alpha$ distance, the decrease of the energy due to the tensor interaction
is enhanced; about 20 MeV around the lowest energy point.
Therefore, it can be concluded that the tensor interaction has a certain effect for the stability of clusterized configurations.

In the present case, this 2p2h tensor effect is already renormalized in the central part
of the effective interaction as in many conventional models, 
and the result gives very large overbinding.  
As a future work, we modify the central part of the two-body and three-body interactions
to reproduce the binding energies of many nuclei including this kind of tensor effect.

\begin{acknowledgments}
Numerical calculation has been performed using the computer facility of 
Yukawa Institute for Theoretical Physics,
Kyoto University. This work was supported by JSPS KAKENHI Grant Number 17K05440.
\end{acknowledgments}

\end{document}